\documentclass[
journal=prl,
doi=true, superscriptaddress,
manuscript=article,layout=twocolumn, reprint, floatfix
]{revtex4-1}

\usepackage[version=3]{mhchem} %

\usepackage{dcolumn}%
\usepackage{bm}%
\usepackage{amsfonts}
\usepackage{bbm}
\usepackage{todonotes}
\usepackage{tikz}
\usepackage{cool}
\usepackage{dirtytalk}
\usepackage{physics}
\usepackage[fleqn]{mathtools}
\usepackage{isomath}
\usepackage{lmodern}
\usepackage[english]{babel}
\usepackage{multirow} 

\makeatletter
\def\bbl@set@language#1{%
  \edef\languagename{%
    \ifnum\escapechar=\expandafter`\string#1\@empty
    \else\string#1\@empty\fi}%
  \@ifundefined{babel@language@alias@\languagename}{}{%
    \edef\languagename{\@nameuse{babel@language@alias@\languagename}}%
  }%
  \select@language{\languagename}%
  \expandafter\ifx\csname date\languagename\endcsname\relax\else
    \if@filesw
      \protected@write\@auxout{}{\string\select@language{\languagename}}%
      \bbl@for\bbl@tempa\BabelContentsFiles{%
        \addtocontents{\bbl@tempa}{\xstring\select@language{\languagename}}}%
      \bbl@usehooks{write}{}%
    \fi
  \fi}
\newcommand{\DeclareLanguageAlias}[2]{%
  \global\@namedef{babel@language@alias@#1}{#2}%
}
\makeatother

\DeclareLanguageAlias{en}{english}

\definecolor{darkgreen}{rgb}{0.0, 0.4, 0.13}

\begin{document}

\title{Efficient Quantum Vibrational Spectroscopy of Water with High-Order Path Integrals: from Bulk to Interfaces}

\author{Sam Shepherd}
\affiliation{Atomistic Simulation Centre, School of Mathematics and Physics, Queen’s University Belfast, Belfast BT7 1NN, Northern Ireland, United Kingdom}

\author{Jinggang Lan}
\affiliation{Department of Chemistry, University of Z\"urich, Z\"urich, Switzerland}

\author{David M. Wilkins}
\email{d.wilkins@qub.ac.uk}
\affiliation{Atomistic Simulation Centre, School of Mathematics and Physics, Queen’s University Belfast, Belfast BT7 1NN, Northern Ireland, United Kingdom}

\author{Venkat Kapil}
\email{vk380@cam.ac.uk}
\affiliation{Yusuf Hamied Department of Chemistry,  University of Cambridge,  Lensfield Road,  Cambridge,  CB2 1EW,UK}

\begin{abstract}
\noindent Vibrational spectroscopy is key for probing the interplay between the structure and dynamics of aqueous systems.
In order to map different regions of experimental spectra to the microscopic structure of a system, it is important to combine them with first-principles atomistic simulations that incorporate the quantum nature of nuclei.
Here, we show that the large cost of calculating quantum vibrational spectra of aqueous systems can be dramatically reduced compared to standard path integral methods by using approximate quantum dynamics based on high-order path integrals.
Together with state-of-the-art machine-learned electronic properties, our approach gives an excellent description of the infrared and Raman spectra of bulk water, but also of 2D correlation and more challenging sum-frequency generation spectra of the water-air interface. 
This paves the way for understanding complex interfaces such as water encapsulated between or in contact with hydrophobic and hydrophilic materials, through robust and inexpensive surface sensitive and multidimensional spectra with first-principles accuracy. \end{abstract}

\maketitle

Vibrational spectroscopies provide a great deal of information about the structure and dynamics of aqueous systems, ranging from pump-probe Infrared (IR) experiments that also give access to rotational dynamics of water molecules~\cite{Bakker2004,Bakker2005,Piatkowski2011} to sum-frequency generation (SFG) spectroscopy which is sensitive to ``dangling'' O--H bonds at the interface between water and a hydrophobic system~\cite{Shen1989a,Shen2006,Roke2012,Nagata2012,Chen2013}, and two-dimensional IR spectroscopy which gave evidence for the angular jump mechanism of hydrogen-bond acceptor exchange~\cite{Laage2006,Laage2008,Laage2008a,Stirnemann2013}. 
However, the main difficulty in interpreting vibrational spectra is that it is not always possible to find an unambiguous relation between experimentally observed spectral features and the microscopic structure of a material. 
Theoretical calculations, particularly using molecular dynamics (MD) simulations, are able to help in making the connection between the two~\cite{McQuarrie2000}, allowing the effects of different types of motif on the spectra to be isolated~\cite{Nagata2012,Medders2016,Marsalek2017}. \\

An accurate description of the interatomic interactions in aqueous systems often requires computationally expensive calculations based on density functional theory (DFT)~\cite{Gillan2016,Marsalek2017} or beyond~\cite{Lan2021}, along with first-principles calculations of the system's electric response properties.
Beyond the inherently quantum-mechanical nature of the potential energy surface (PES) of the system, it is well-established that in systems containing light nuclei, the \textit{motion} on the PES is also affected by nuclear quantum effects (NQEs) such as zero-point energy and tunnelling~\cite{Kuharski1984,Wallqvist1985,Miller2005,Paesani2009,Ceriotti2016,Wilkins2017}. 
These effects can be accounted for using path integral methods such as (thermostatted) ring polymer molecular dynamics ((T)RPMD)~\cite{Craig2004,Craig2005a,Habershon2013,Rossi2014} and (partially adiabatic) centroid molecular dynamics ((PA)CMD)~\cite{Cao1993,Cao1994,Jang1999,Hone2006}, which generally add an extra level of computational expense due to the large number of replicas of the system required.
These factors combine to make the theoretical calculation of vibrational spectra of water a difficult undertaking, often requiring compromises to be made with the level of theory used, and thus with the accuracy. \\

In a previous paper, some of the current authors addressed these challenges by using high-quality machine-learning (ML) predictions of the PES~\cite{behler_generalized_2007,kapil_high_2016,Cheng2019} and the electric response properties~\cite{Grisafi2018,Grisafi2019} of aqueous systems to circumvent the need for expensive quantum-mechanical calculations~\cite{kapil_inexpensive_2020}.
These predictions were combined with coloured-noise generalized Langevin equation (GLE) thermostats~\cite{Ceriotti2012}, which mimic nuclear quantum fluctuations and allow their effect on the dynamics to be approximated~\cite{Rossi2018, kapil_inexpensive_2020} at a much lower cost than standard path integral methods. 
While the action of the thermostat perturbs the centroid dynamics, and thus the vibrational spectrum, it is possible to correct for this perturbation using a deconvolution procedure described in Ref.~\onlinecite{Rossi2018}.
The spectra found using this approach are in good agreement with the results of TRPMD and PA-CMD, but with a reduction in computational cost by a factor of 4-5~\cite{kapil_inexpensive_2020}.
However, it proved unable to recover subtle low-intensity features in Raman spectra, particularly when the noise due to deconvolution is comparable to the intensities of the modes.
Furthermore, the deconvolution algorithm used within this scheme is limited to recovering positive-definite signals~\cite{daube-witherspoon_iterative_1986, archer_bayesianregularization_1995}, precluding its use for multidimensional and nonlinear spectra, which may have negative intensities. \\

In this letter, we develop an alternative method based on higher-order path integrals, which allows for an accurate and inexpensive estimation of vibrational spectra without the limitations of the GLE thermostat scheme.
In tandem with state-of-the-art machine-learned electric response property surfaces, we show that
our approach gives an excellent description not only of the IR and Raman spectra of bulk water, but also of 2D correlation and more challenging SFG spectra of water at interfaces. 
This results in a robust and computationally efficient scheme that can be applied to model a wide range of spectroscopies and dynamical properties. \\

\begin{figure*}[t]
\begin{center}
\includegraphics[width=\textwidth]{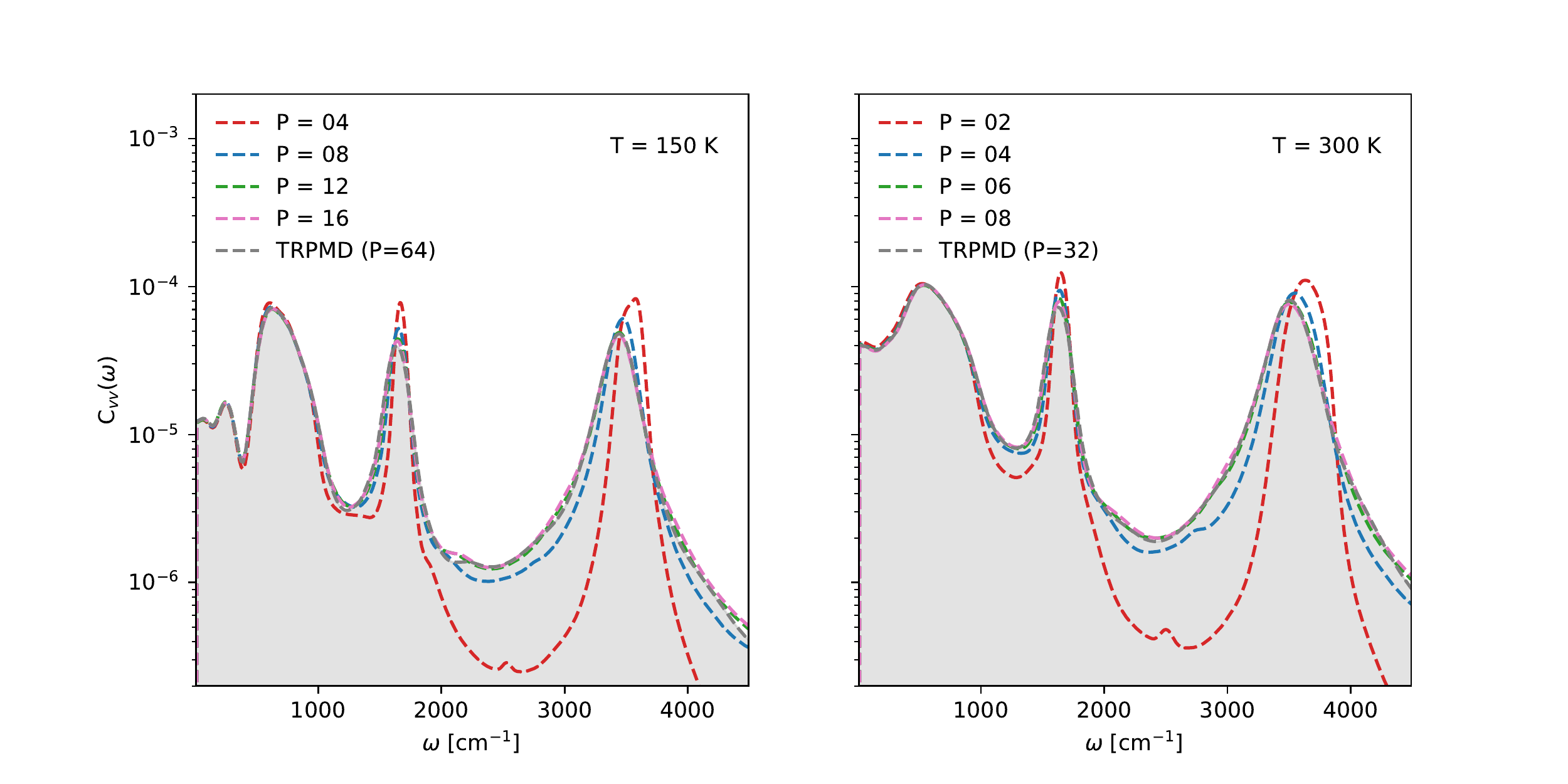}
\end{center}
\caption{Vibrational  density of states (VDOS) for Takasha-Imada TRPMD simulations of hexagonal ice at 150 K (left) and liquid water at 300 K (right), using up to 16 beads (for red, blue, green and pink dashed lines) with the correction of Eq.~\eqref{eq:redshift} applied in all cases.
For reference, the grey shading indicates the VDOS calculated using standard second-order TRPMD. \label{fig:qtip4pf}}
\end{figure*}

The use of high-order splittings of the Boltzmann operator, such as the Suzuki-Chin~\cite{Suzuki1995} and Takashi-Imada (TI)~\cite{Takahashi1984} schemes, are a popular method of reducing the computational cost of path integral calculations~\cite{markland_nuclear_2018}. %
These approaches converge the quantum Boltzmann distribution faster with the number of replicas $P$, compared to the standard approach based on the second-order splitting~\cite{trotter_product_1959}, and can be applied to general condensed phase systems in a computationally efficient manner~\cite{Kapil2016}.
The terms in higher-order ring polymer Hamiltonians that accelerate the convergence of quantum statistics also perturb the dynamics of the centroid, leading to incorrect quantum dynamics for finite $P$. 
For this reason, higher-order path integral methods have hitherto been considered unsuitable for computing quantum dynamical properties~\cite{Perez2009, Kapil2016}. \\

Here we present a simple fix that can be used to correct for this perturbation within the Takashi-Imada scheme, allowing the use of higher-order path integral calculations for spectroscopic properties.
We begin by noting that for a one-dimensional harmonic oscillator with natural frequency $\omega_{0}$, the fourth-order TI Hamiltonian in mass-scaled coordinates is,
\begin{equation}
H^{(4)}_P = \sum_{j=1}^{P}\frac{1}{2}\abs{\tilde{p}^{(j)}}^2 + \frac{1}{2}  \omega_P^{2}  \left[\tilde{q}^{(j)} - \tilde{q}^{(j+1)}\right]^2 + \frac{1}{2} \omega_{\text{eff}}^{2} \abs{\tilde{q}^{(j)}}^2,   
\end{equation}
where $P$ is the number of replicas, $\tilde{p}^{(j)} = p^{(j)}/\sqrt{m}$ is the mass-scaled momentum of the $j^{\rm th}$ replica, $\tilde{q}^{(j)} = \sqrt{m} q^{(j)}$ its mass-scaled coordinate, $\omega_{P} = P / \beta \hbar$, and $\omega_{\text{eff}}^{2} = \omega_{0}^{2} + \frac{\omega_{0}^{4}}{12 \omega_{P}^{2}}$
is the effective harmonic frequency experienced by each replica. This suggests that within the (T)RPMD or (PA)CMD scheme, which are exact for harmonic potentials, the main effect of TI dynamics on a system's vibrational density of states is a blue shift of frequencies from $\omega$ to $\omega_{\text{eff}}$. 
For a harmonic oscillator, P{\'e}rez and Tuckerman suggested altering the mass to correct for this shift, but for general potentials considered the TI scheme unsuitable for approximating quantum dynamics~\cite{Perez2009}.
However, when the Hamiltonian is written in mass-scaled coordinates the dependence of frequency on the system's mass is eliminated, suggesting that if a system is assumed to be a collection of harmonic oscillators we can systematically apply a red shift, obtaining a corrected spectrum $I_{0}(\omega)$ from (T)RPMD or (PA)CMD spectrum of the TI Hamiltonian (referred to as TI-(T)RPMD and TI-(PA)CMD respectively) $I_{\text{TI}}(\omega)$ by using the relation $I_{0}(\omega) = I_{\text{TI}}(\omega_{\text{TI}})$ with,
\begin{equation}
\omega = \omega_{P} \sqrt{-6 + 6 \sqrt{1 + \frac{\omega_{\text{TI}}^{2}}{3\omega_{P}^{2}}}}.
\label{eq:redshift}
\end{equation}
In practice this red-shift amounts tp a simple rescaling of the frequency axis.
To correct for the line widths of the spectra, we ensure that the differential area under the curve is invariant to the transformation of the frequencies. \\

\begin{figure*}[tbh]
\begin{center}
\includegraphics[width=\textwidth]{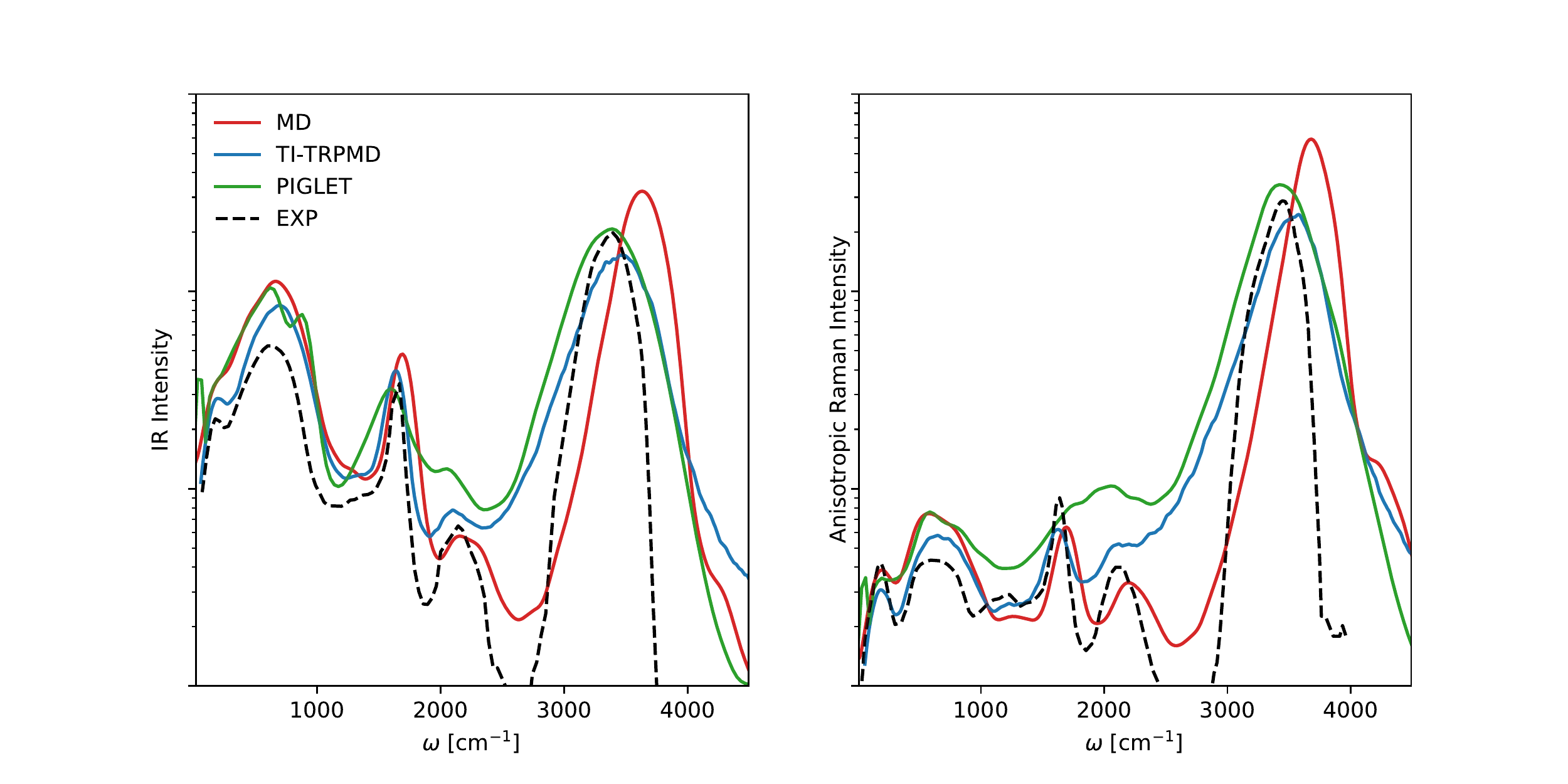}
\end{center}
\caption{\label{fig:IRRaman} IR (left) and anisotropic Raman (right) spectra of liquid water at 300\,K, using classical MD (red), Takashi-Imada TRPMD (blue) and PIGLET with a deconvolution scheme applied (green). These  results are compared to the experimental spectra (black dashed lines) of Refs.~\cite{brooker_raman_1989} and \cite{bertie_infrared_1996} for IR and Raman respectively.}
\end{figure*}

In order to study the performance of the proposed correction, we calculate the vibrational density of states (VDOS) of hexagonal ice at 150\,K and of liquid water at 300\,K using the q-TIP4P/F potential~\cite{habershon_competing_2009}.
The VDOS is given by the system's velocity autocorrelation function, with the blue-shift of the TI-TRPMD results corrected using Eq.~\eqref{eq:redshift}.
Fig.~\ref{fig:qtip4pf} compares the VDOS computed using TI-TRPMD with varying numbers of ring-polymer replicas to the converged TRPMD result using 32 replicas at 300\,K and 64 replicas at 150\,K.
The VDOS from higher-order path integrals converge much more quickly with respect to number of replicas than do those from TRPMD, with 12 beads required for ice and 6 beads for water. Furthermore, the shapes and positions of the vibrational bands in the TI-TRPMD VDOS are almost indistinguishable from those of the reference TRPMD density of states.
These results indicate that, despite being derived in the harmonic limit, the correction of TI-TRPMD spectra with a simple red-shift is extremely accurate even for a highly anharmonic liquid water.
Although the TI scheme requires the calculation of higher-order forces, finite difference and multiple timestep methods mean that these calculations add little extra computational expense~\cite{kapil_high_2016}. The VDOS calculated using TI-TRPMD are 4-5 times less expensive than standard TRPMD, with no appreciable loss in accuracy. \\

\begin{figure*}[tbh]
\begin{center}
\includegraphics[width=0.9\textwidth]{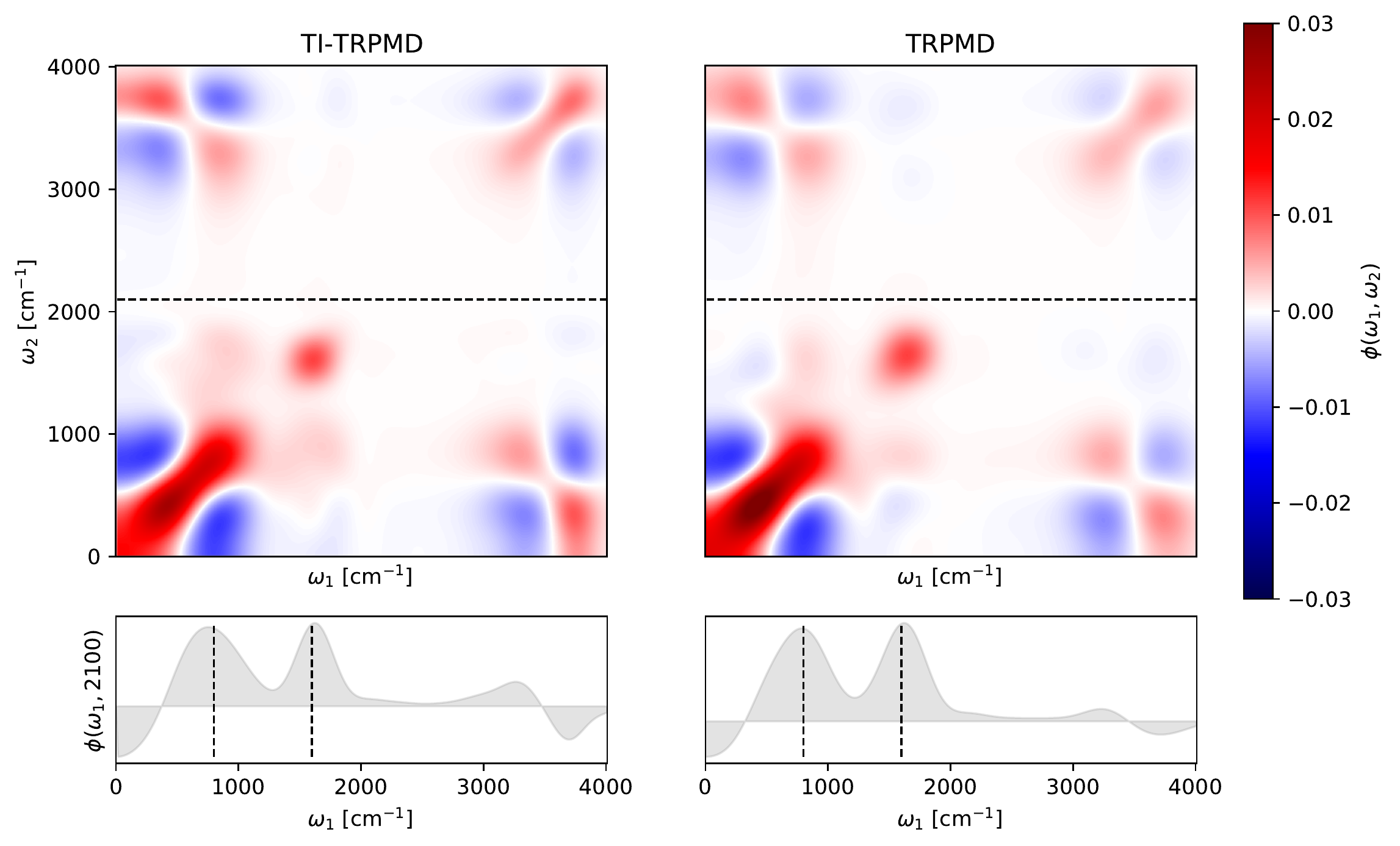}
\end{center}
\caption{\label{fig:2D} 2D vibrational Raman correlation spectrum of water modelled using a machine learning potential trained at revPBE0+D3 level of theory calculated using Takashi-Imada (left) and standard (right) TRPMD. Positive values indicate a temporal correlation between two vibrational modes, while negative values indicate an anti-correlation. The bottoms panels show the cut of the 2D correlation spectrum at the combination band at 2100 $\text{cm}^{-1}$.
}
\end{figure*}

As a more stringent test of our methodology, we compute vibrational spectra with first-principles accuracy, combining molecular dynamics simulations using neural network potentials~\cite{cheng_ab_2019} trained on \textit{ab initio} data with symmetry-adapted Gaussian process regression (SA-GPR)~\cite{Grisafi2018} models for the dielectric response tensors, as proposed by some of the authors  in Ref.~\onlinecite{kapil_inexpensive_2020}.
Fig.~\ref{fig:IRRaman} shows the IR and anisotropic Raman spectra of liquid water at 300\,K using classical molecular dynamics, TI-TRPMD, and for comparison, the recently proposed accelerated method~\cite{kapil_inexpensive_2020} using the PIGLET thermostat.
In agreement with our previous work~\cite{kapil_accurate_2016}, the PIGLET method is able to accurately capture the red shift of the O--H stretching frequency leading to a good agreement of the line position with the IR and Raman experimental results, as shown in Fig. ~\ref{fig:IRRaman}. 
However, the IR bending mode is broadened, and the low frequency IR region contains noise originating from relatively poor statistics for long time collective motion. 
This problem is aggravated in the case of the anisotropic Raman spectrum where the intensities of the high and low frequency modes differs by about a factor of 50.
The noise induced by the deconvolution algorithm is comparable to the weak signal of the modes below 2000\,cm$^{-1}$ and thus washes out the low frequency part of the spectrum. 
On the other hand, the correction to the TI scheme remains independent of the band intensities as it only requires a rescaling of the frequency scale. 
The TI-TRPMD spectra remain in good agreement with the reference TRPMD~\cite{marsalek_quantum_2017} and the experimental spectra across the \textit{entire} frequency range. 
The improvement with respect to the PIGLET approach is especially apparent for the Raman spectra where both line positions and shapes of the low and high frequency modes are accurately captured at a comparable computational cost.
The small disagreement of the stretching line position with the experimental results is due to the inherent limitations of the TRPMD scheme~\cite{marsalek_quantum_2017}, rather than of the TI scheme. 
The generality of our approach means that TI corrections to PACMD or the recently developed quasi-centroid molecular dynamics (QCMD)~\cite{trenins_path-integral_2019} can be used to alleviate this disagreement, and enable accurate predictions of vibrational spectra of a diverse range of systems. \\

To understand if the TI correction scheme can also describe coupling between vibrational modes, we estimate the the 2D correlation spectrum $\phi(\omega_{1},\omega_{2})$.
For a simulation split into $m$ blocks of time $T$ the spectrum is defined,
\begin{equation}
\phi(\omega_1,\omega_2) = \frac{1}{m-1}\sum_{j=1}^{m} \left(I_j(\omega_1) - \bar{I}(\omega_1)\right) \left(I_j(\omega_2) - \bar{I}(\omega_1)\right),
\end{equation}
where $I_{j}(\omega)$ is the vibrational density of the system in the $j^{\rm th}$ block and 
$\bar{I}(\omega)$ is the average vibrational density over the blocks~\cite{Raimbault2019}.
The diagonal peaks in this spectrum indicate how fast the intensity changes over the time interval, while off-diagonal peaks indicate simultaneous changes at different frequencies, which can arise from coupling between modes.
While the 2D correlation spectrum does not give information into temporal coherence of these modes or into vibrational relaxation due to coupling, it remains a useful and easy-to-compute tool for understanding mode coupling and combination bands. 
These spectra have been used to understand the origin of combination bands in aqueous systems~\cite{morawietz_interplay_2018} and to study coupling between intra- and intermolecular modes across different polymorphs of molecular crystals~\cite{Raimbault2019}. \\

Fig.~\ref{fig:2D} shows the 2D Raman correlation spectrum of liquid water at 300\,K.
Although the TI correction is derived in the harmonic limit, it successfully reproduces the correlations between vibrational intensities in excellent agreement with standard TRPMD. Further, we find that the TI scheme also gives a good description of coupling between modes, as evidenced by the slice of the 2D correlation spectrum at 2100\,cm$^{-1}$ for the combination band of the librational (800\,cm$^{-1}$) and bending mode (1600\,cm$^{-1}$). 
These results indicate that the TI scheme is able to capture correlations between modes, an encouraging result for future studies of multidimensional spectroscopy, which is more challenging to calculate when including nuclear quantum effects due to the prohibitive computational cost~\cite{Hamm2009}.
Recent work suggests that the symmetrized Kubo-transform of these time correlation function can be approximated using Matsubara dynamics~\cite{Hele2015, Hele2015a}.
This means that methods such as TRPMD and PACMD can be used for practical evaluations of these functions, and thus of multidimensional spectra~\cite{tong_two-dimensional_2020}, and our TI correction could be an efficient route towards its calculation. \\

Having shown the ability of higher-order path integrals to efficiently compute vibrational spectra of bulk aqueous systems, we finally apply it to sum-frequency generation (SFG) spectroscopy, a powerful tool to study
the structure and dynamics of surface phenomena~\cite{shen_optical_1989,roke_nonlinear_2012}. 
It has been used to investigate the interface of water with both hydrophobic~\cite{Gan2006,Samson2014} and hydrophilic~\cite{Miranda1999,Rehl2018} substances, and the complexity of SFG measurements means that atomistic simulations are important for a full understanding.
The calculation of this spectrum requires the cross-correlation function between polarizations and polarizabilities, which only gives a nonzero contribution in non-centrosymmetric regions.
Since the number of molecules at the interface are fewer than those in bulk, calculating a converged SFG spectrum usually requires much more simulation time than other vibrational spectra. \\

\begin{figure}[tbh]
\begin{center}
\includegraphics[width=0.5\textwidth]{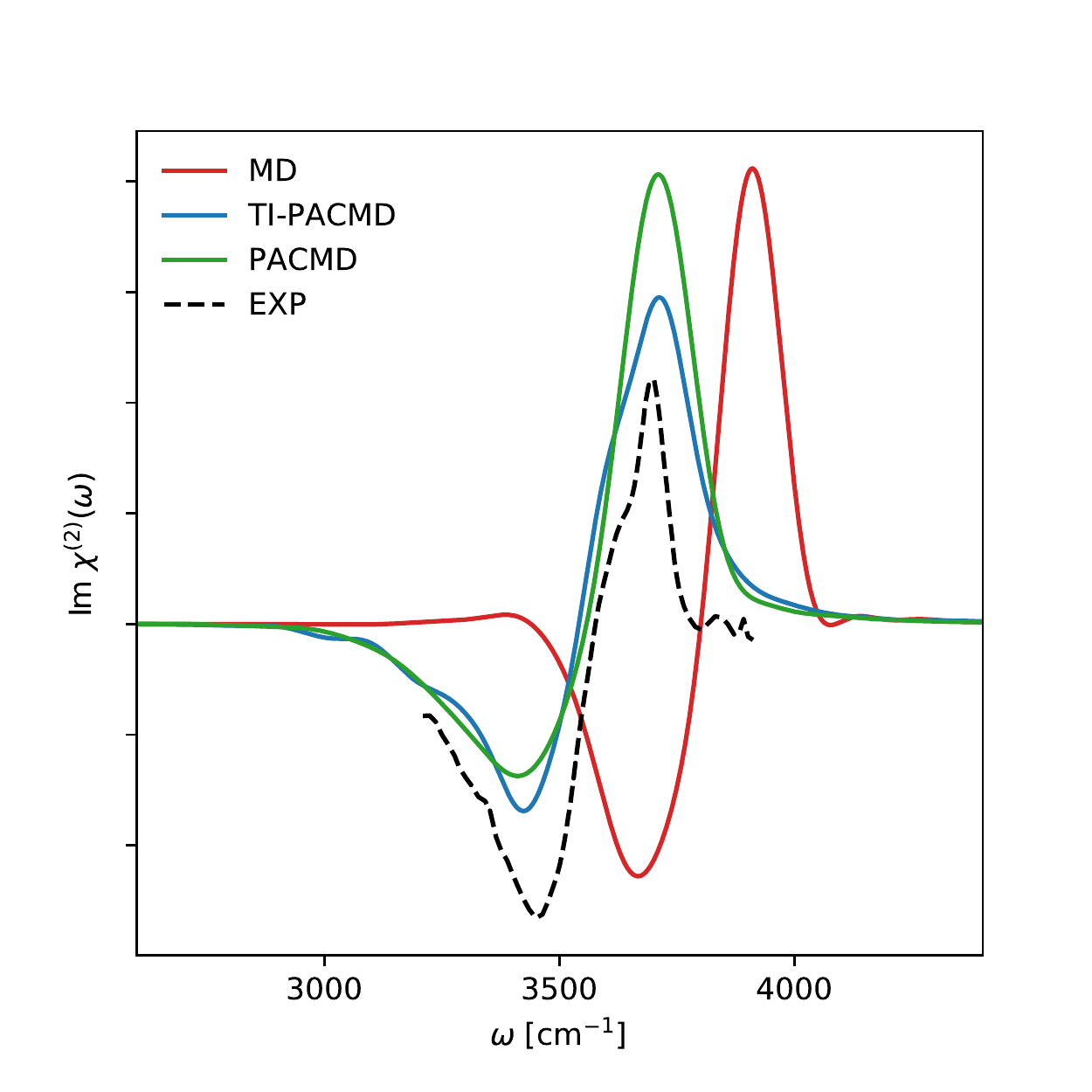}
\end{center}
\caption{\label{fig:SFG} Sum-frequency generation (SFG) spectrum of the water-vacuum interface at 300\,K, approximated using the surface-sensitive velocity autocorrelation function~\cite{ohto_toward_2015} with classical MD (red), Takashi-Imada PACMD (pink) and PACMD (blue). The black dashed line shows experimental results~\cite{nihonyanagi_accurate_2015}.
}
\end{figure}

Medders and Paesani have shown that NQEs are critical for
accurately assigning spectral features to the correct local motif at the air-water interface~\cite{Medders2016}. 
Unfortunately, the large cost of first-principles approaches and the statistics required to obtain a converged SFG spectrum means that a fully \textit{ab initio} SFG spectrum for classical nuclei is already computationally prohibitive.
For non-dissociative aqueous systems, Ohto \textit{et al.} have shown that the classical SFG spectrum can be approximated by computing a surface-sensitive velocity autocorrelation function (SSVAF)~\cite{ohto_toward_2015,ohto_accessing_2019} which converges much faster than the traditional polarization-polarizability cross correlation. 
Fig.~\ref{fig:SFG} shows the quantum SFG spectrum of the air-water interface calculated using a (TI-)PACMD estimate of the SSVAF, with a ML potential trained at the revPBE0+D3 level for interfacial systems~\cite{schran_committee_2020}.
Inclusion of nuclear quantum effects using PACMD leads to a large improvement, compared to classical MD, in the line positions of the stretching bands of the hydrogen-bonded O--H and dangling O--H bonds~\cite{Marsalek2017}.
Furthermore, the TI correction gives a very good agreement with the standard PACMD spectrum at a 4-5 times lower computational cost than standard path-integral approaches.
This is a challenging test for our scheme as the dangling O--H bonds at the water-air interface exhibit large amplitude curvilinear motion, which causes systematic errors in a range of approximate methods based on a harmonic approximation~\cite{rossi_fine_2017, willatt_approximating_2018, kapil_assessment_2019}.
The efficiency and accuracy of our scheme alleviates the need to apply an \textit{ad hoc} red shift to classical SFG spectra to account for nuclear quantum effects~\cite{ohto_accessing_2019}, and enables predictive modelling of surface sensitive spectroscopy at first-principles accuracy. \\

In summary, we have shown that the unphysical blue shift inherent in vibrational spectra calculated using the TI path integral method can be corrected using a simple rescaling of frequencies. 
This results in a robust approach that is 4-5 times less costly than standard methods of simulating quantum nuclear effects, and is free from the deficiencies of previous developments in acceleration techniques~\cite{kapil_inexpensive_2020} for quantum dynamics. 
Our approach is highly accurate for anharmonic systems exhibiting quantum nuclear motion such as liquid water and its interface with vacuum, allowing accurate modeling of spectroscopies such as IR, Raman, and SFG.
Furthermore, it shows great promise for predicting 2D pump-probe IR experiments
as it captures correlations between vibrational modes, and also for modeling of SFG experiments to study water dissociation at interfaces by directly computing the polarization-polarizability cross-correlation function using ML predictions of dielectric responses.
These results pave the way for future work on understanding of more complex systems such as water in confinement, and interfaces between water and hydrophobic materials or metals by means of quantum-mechanical studies. \\

Although we have focused on vibrational spectra, we note that high-order path integrals could be used for accelerated calculations of other dynamical properties. 
In the zero-frequency limit, the TI correction of Eq.~\eqref{eq:redshift} has no effect, as evidenced by the good description of the low frequency region of velocity-velocity autocorrelation functions in Fig.~\ref{fig:qtip4pf} by the TI scheme.
This suggests that high-order dynamics will allow the direct calculation of properties such as diffusion and thermal transport coefficients and rate constants, which are calculated as $\omega\rightarrow 0$ limits of the Fourier transform of Green-Kubo time correlation functions.
In much the same way as recent developments in accelerated methods have made feasible static path integral calculations~\cite{markland_nuclear_2018}, we expect that methods such as that developed in Ref.~\cite{kapil_inexpensive_2020} and in this work will bring the inclusion of quantum effects in the calculation of dynamical properties into the mainstream.

\section*{Acknowledgement}
D.M.W. thanks Queen's University Belfast (QUB) for startup funding.
V.K. acknowledges support by  Swiss National Science Foundation (SNSF).
This work was supported by a grant from the Swiss National Supercomputing Centre (CSCS) under Project ID s1000, and by computer time from QUB. We are also grateful to the UK Materials and Molecular Modelling Hub for computational resorces, which is partially funded by EPSRC (EP/P020194/1 and
EP/T022213/1). The authors thanks Christoph Schran, Yair Litman, Michele Ceriotti and Stuart Althorpe for insightful discussions and useful comments on the manuscript.  

\end{document}